# MEASUREMENTS OF THERMOPHYSICAL PROPERTY OF THIN FILMS BY LIGHT PULSE HEATING THERMOREFLECTANCE METHODS

*Tetsuya Baba, Kazuko Ishikawa, Takashi Yagi, Naoyuki Taketoshi*

National Metrology Institute of Japan,
National Institute of Advanced Industrial Science and Technology
AIST Tsukuba Central 3, Umezono 1-1-1, Tsukuba, Ibaraki, 305-8563, Japan

*Kimiaki Tamano, Tetsuro Ohtsuka, Hiroshi Watanabe, Yuzo Shigesato*

Department of Chemistry School of Science & Engineering Aoyama Gakuin University,
Fuchinobe, Sagamihara, Kanagawa, 229-8558 Japan

## ABSTRACT

Thermoreflectance methods by picosecond light pulse heating and by nanosecond light pulse heating have been developed under the same configuration as the laser flash method by National Metrology Institute of JAPAN, AIST. Using these light pulse heating thermoreflectance methods, the thermal diffusivity of each layer of the multilayered thin films and the boundary thermal resistance between the layers can be determined from the observed transient temperature curves based on the response function method. Various thin films as the transparent conductive films used for flat panel displays, hard coating films and multilayered films of next generation phase-change optical disk have been measured by these methods.

## 1. INTRODUCTION

Phase-change optical disk media (DVD Rewritable), high-density integrated circuit and flat panel display are comprised of several nanometers to several 100 nanometers thick of thin films. To know how the heat flows and what distribution of temperature is caused when they are used, information of the thermophysical properties of thin films and the boundary thermal resistance between thin films is required [1, 2].

Picosecond thermoreflectance method was developed to measure thermal diffusivity of subnanometer thick thin films by Paddock and Eesley [3]. The optical reflection intensity of the temperature detection light is detected by photodiode. Since reflectivity of material surface changes dependent on the surface temperature, the change of specimen front face temperature can be observed by the change of reflected light amplitude. This temperature measurement method with the temperature change of such a reflectivity is called as thermoreflectance method [4]. Thermal diffusivity of submicrometer thin films perpendicular to the surface was calculated from the cooling rate of the surface temperature and the penetration depth of the heating light.

National Metrology Institute of JAPAN, AIST has succeeded in developing the thermoreflectance methods by picosecond / nanosecond light pulse heating [5, 6, 7] and realized to measure the thermal diffusivity of metallic thin films from several 10 nm to several micrometers thick on transparent substrate in thickness direction under the configuration of rear face heating / front face detection picosecond thermoreflectance method [1, 8-11].

Since the geometrical configuration of this method is the same as the laser flash method which is the standard measurement technique for the thermal diffusivity of bulk materials [12, 13], thermal diffusivity value can be calculated reliably from heat diffusion time across well-defined length of the film thickness under one-dimensional heat flow [8, 9].

## 2. HIGH SPEEDLIGHT PULSE HEATING THERMOREFLECTANCE METHODS

### 2.1. Front face heating / front face detection

It was not easy by conventional measurement technique to determine the thermal conductivity and the thermal diffusivity of thickness direction of thin films of less than one micrometer thick. In order to solve this problem, picosecond thermoreflectance method was developed to observe the temperature changes of thin film front face by heat diffusion to the inside [3]. The optical reflection intensity of the temperature detection light is detected by photodiode based on the thermoreflectance method.

In this thin film thermal diffusivity measurement system by a picosecond thermoreflectance method, the laser beam emitted from a mode lock titanium sapphire laser and is divided into transmitted beam and reflected beam by a quartz plate. About 90 % is used for pulse heating and the other about 10 % is used for temperature detection to measure the temperature changes of the thin film front face. Light travels 0.3 mm in one picosecond. By adjusting distance to a specimen after it was divided, the time difference that a heating light and detection light arrive at the specimen front face can be controlled. The response time of the thermoreflectance method is much faster than thermocouples, resistance temperature sensor or radiation thermometers. According to the pump probe method, ultra fast thermometry is possible only limited by time duration of the pulses. On the other hand, it is a weak point that the sensibility of temperature detection is low.

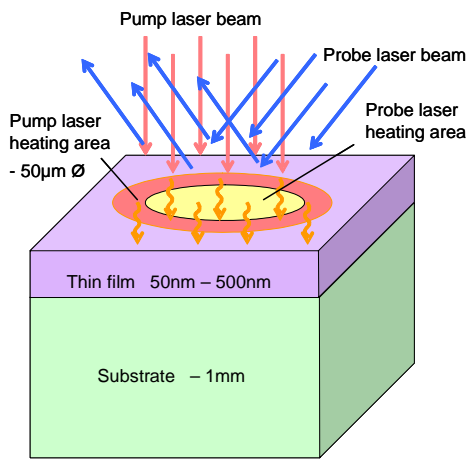

Figure 1  Heating area in front face heating / front face detection picosecond thermoreflectance method

Figure 2 shows the result which observed the change of the front face temperature by the picosecond thermoreflectance method about three kinds of aluminium thin films of different thickness synthesized on a glass substrate [6].

For the specimen of 500 nm thick, the heat has not arrived at the substrate within 120 ps after pulse heating in and the front face temperature change represented by red line corresponds to internal heat diffusion of the aluminium thin film.

On the other hand, for the specimen of 100 nm thick, the temperature change speed decreases around 30 ps after pulse heating as show by black line and deviates from the temperature change of the 500 nm thick specimen.

Because the thermal diffusivity of glass substrate is much smaller than the thermal diffusivity of aluminium thin film, the heat diffusion to the substrate is suppressed when the heat arrives at the interface between the thin film and the substrate.

As shown by blue line in figure 2, for the specimen of 50 nm thick, the temperature change only for the inside of the thin film cannot be observed because of heat diffusion to the substrate just after pulse heating.

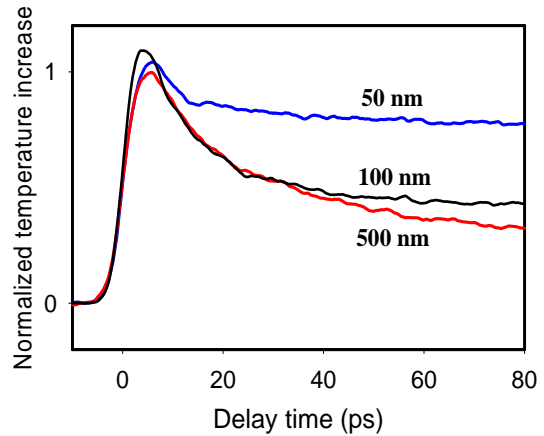

Figure 2  Thermoreflectance signals of three kinds of different thickness of aluminium thin film that synthesized on a glass substrate

Area of a diameter of several 10 μm on thin film front face is heated by the picosecond laser beam and the same position is irradiated by the probe beam. Then, the history of front face temperature is observed by the conventional thermoreflectance method.

In this method, the thermal diffusivity can be calculated from the cooling rate after pulse heating. However, it is not easy to make quantitative and reproducible measurement because the cooling rate changes sensitively dependent on the condition of thin film front face.

**2.2. Rear face heating / front face detection**

National Metrology Institute of JAPAN, AIST has developed rear face heating / front face detection picosecond light heating thermoreflectance methods which are evolution of the conventional laser flash method and the picosecond thermoreflectance method as shown in figure3 [8-11]. Figure 4 shows the block diagram of a measurement system. This configuration is essentially equivalent to the laser flash method which is the standard measurement method to measure the thermal diffusivity of bulk materials. The thermal diffusivity of the thin films can be calculated with small uncertainty from the thickness of a thin film and the heat diffusion time across a thin film.

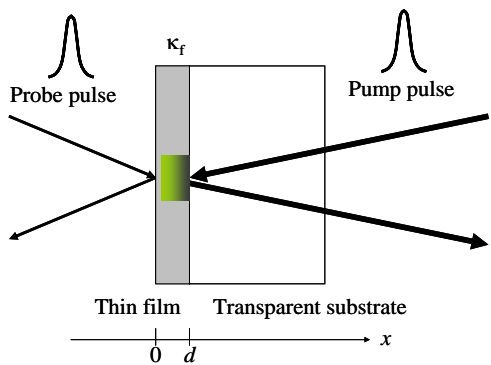

Figure 3   Principle of rear face heating / front face detection picosecond light heating thermoreflectance method

The temperature detection beam is focused to a spot diameter of about 50 μm at the specimen front face just opposite to the heating light focus position. The reflected light intensity of the temperature detection light is in proportion to the change of specimen front face temperature and the change of the reflected light intensity is detected by photodiode. The temperature coefficient of reflectivity for normal metal such as aluminium is small with $10^{-5}/K$ order. Since the transient temperature rise of the specimen front face after picosecond pulse heating is smaller than a few degrees, the thermoreflectance signal is much smaller than the offset level of the reflected light. Such a small signal can be measured by lock-in detection at modulation frequency of heating light by an acoustic optic modulator.

Figure 5 shows the temperature history curves of an aluminium single-layered thin film of 100 nm thick and a molybdenum single-layered thin film of 100 nm measured by the picosecond light pulse heating thermoreflectance method [6]. Both films were synthesized on a Pyrex glass substrate by magnetron DC sputtering method. Here, film thicknesses are nominal values.

Since thermoreflectance signals are similar to those observed by the laser flash method for bulk specimen suggest, the heat energy transport of these metallic thin films of about 100 nm thick at room temperature in time scale of several 100 ps can be expressed by the classic thermal diffusion equation .

### 2.3. Nanosecond light pulse heating thermoreflectance method

Initially, the electrical delay method was developed to expand the observation time of the picosecond light pulse heating thermoreflectance method longer for measurements of thicker films [11]. Since the pulse duration can be longer than picosecond and the repetition period of pulses is flexible, nanosecond pulse laser can be used for light pulse heating thermoreflectance method [1, 14]. For the pump pulse, pulse duration is 2 ns and pulse interval is 20 μs with intensity modulation of 1 kHz by an acoustic optical modulator. Typical size and shape of the specimen are a disk of 10 mm in diameter or a square of 10 mm on each side. The specimen is irradiated from bottom to the rear face of the specimen by the heating beam and the reflected light of the probe beam is detected by a photodiode. The thermoreflectance signals are detected using a lock-in amplifier. In this system it is possible to measure the thermal diffusivity of thin films with thickness up to several micrometers.

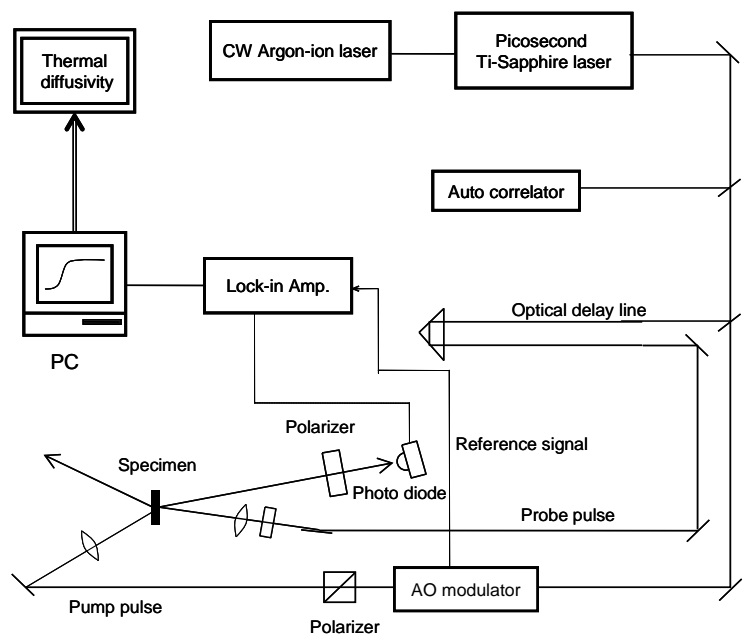

Figure 4   Block diagram of the thermoreflectance method thin film thermophysical property measuring system by picosecond pulsed laser heating

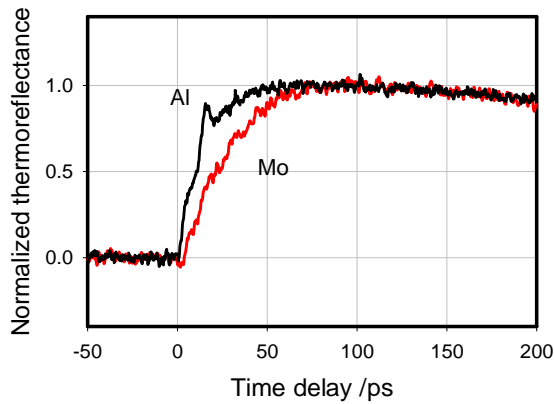

Figure 5   Thermoreflectance signals of aluminium thin film and molybdenum thin film of 100 nm thick by rear face heating / front face detection picosecond thermoreflectance method

### 3. MEASREMENTS OF THIN FILMS FOR INDUSTRIAL USE

#### 3.1. Transparent conductive films

The thermal diffusivity of tin doped indium oxide (ITO) thin films with the thickness of 200 nm and 300 nm have been measured using the nanosecond light pulse heating thermoreflectance method [15]. As shown in figure 6, ITO films between molybdenum (Mo) films of 70 nm thick were prepared on fused silica substrate by DC magnetron sputtering using ITO and Mo multi-targets because the wavelengths of pulse lasers used in this study are 1064 nm as pump beam and 830 nm as probe beam where ITO is transparent. The thermal diffusivity measurements of three-layered films were carried out using the nanosecond light pulse heating thermoreflectance system. During deposition of Mo films, total gas pressure was maintained at 1.0 Pa. On the other hands, ITO layers were deposited under various total gas pressure between 0.5 and 3.0 Pa. Such a Mo/ITO/Mo layered structure was fabricated without exposure to the atmosphere between each deposition.

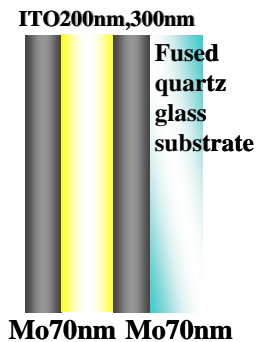

Figure 6   Tin doped indium oxide (ITO) thin films between molybdenum (Mo) films deposited on fused silica substrate

Analysis of heat diffusion across three layer thin films on substrate must be considered in order to calculate the thermal diffusivity of the thin film between metal thin films. It is also necessary to know the boundary thermal resistance between the layers as well as the thermophysical properties of each layer. The measured temperature history curve for the three-layered films were analyzed by the response function method which is general technique to analyze heat diffusion across a multilayer films developed by AIST [1, 7, 16-18].

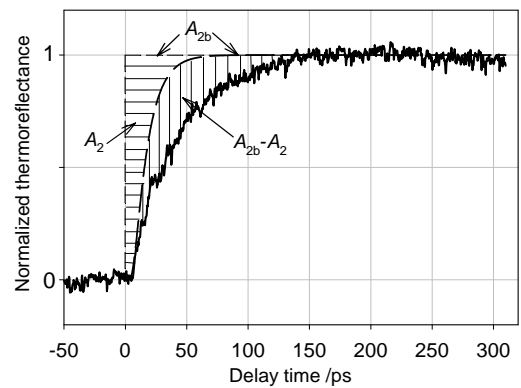

Figure 7   Area surrounded by the maximum temperature rise line and the temperature response at the specimen rear face after the pulse heating

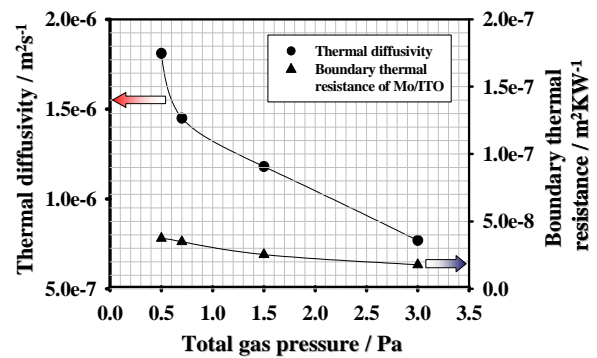

Figure 8   Thermal diffusivity of ITO thin films and boundary thermal resistance between ITO and Mo dependent on total gas pressure

The areal heat diffusion time is defined as the area surrounded by the horizontal line at the height of the maximum temperature rise and the transient temperature response curve at the rear face after pulse heating as shown in figure 7 [1]. Thermal diffusivity value of the thin unknown thin film and the boundary thermal resistance can be calculated from the observed areal heat diffusion time based on the response function method assuming thickness of each film, thermal diffusivity and volume heat capacity of molybdenum thin films are known. The thermal diffusivity of the ITO films decreased as the total gas pressure increased as shown in figure 8 [15].

### 3.2. Hard coating films

Titanium Nitride (TiN) thin films are used as hard coating films. TiN thin films with the thickness from 200 nm to 800 nm were synthesized on glass substrate by reactive RF magnetron sputtering using Ti target and Ar-N2 plasma. Thermal diffusivity measurements of these films were carried out using the nanosecond light pulse heating thermoreflectance system [19]. As shown in figure 9, the thermal diffusivity of TiN film changes dependent on the ratio of N over Ti which is controlled by $N_2$ gas flow ratio in Ar-$N_2$. There is good linear relation between the thermal diffusivity and the electric conductivity as shown in figure 10. The thermal conductivity was calculated using the specific heat capacity and the density of bulk TiN. Then, Lorenz number of TiN thin film was calculated assuming Wiedemann-Franz law. As electrical conductivity increases, Lorenz number converges to Sommerfeld value based on the free electron model. This result means that the major heat carriers are phonons for lower electrical conductive TiN and electrons for higher electrical conductive TiN [19].

### 33. Multilayered films of the next generation phase change optical disk

Dependence of the thermal diffusivity of $Ge_2Sb_2Te_5$ thin films, used for optical recording media, on phase change has been investigated using the nanosecond light pulse heating thermoreflectance system [20]. Two kinds of thin films are prepared; one is $Ge_2Sb_2Te_5$ single layered thin film and the other is Mo/$Ge_2Sb_2Te_5$/Mo three-layered thin film, where 70 nm thick Mo films act as the absorbing layer for the heating beam and the reflection layer for the temperature detection beam. Films, with which the thickness of $Ge_2Sb_2Te_5$ is from 100 nm to 400 nm, were deposited on non-alkali glass substrates by RF magnetron sputtering from $Ge_2Sb_2Te_5$ and Mo targets. It is observed by X-ray diffraction study that the as-deposited films show amorphous structure and it is transformed to FCC crystal phase by heat treatment of 5 min at 573 K. The thermal diffusivity of $Ge_2Sb_2Te_5$ amorphous thin film is determined to be $2.5 \times 10^{-7}$ m$^2$/s and the thermal diffusivity of the crystal film is $4.8 \times 10^{-7}$ m$^2$/s from the temperature history curves shown in figure 11 observed by the nanosecond light pulse heating thermoreflectance system [20].

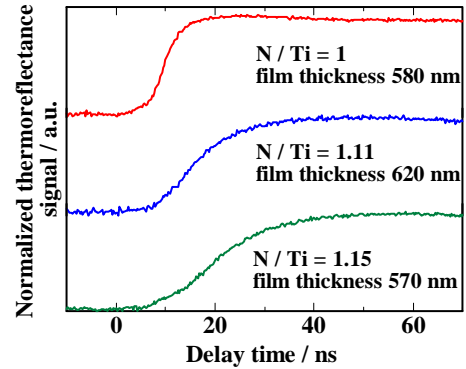

Figure 9   Thermoreflectance signals of TiN thin films of different ratio of N over Ti

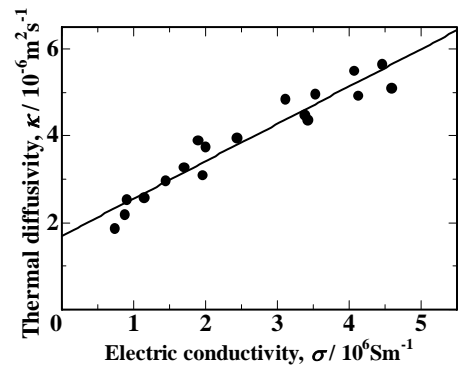

Figure 10   Relationship between thermal diffusivity and electrical conductivity of TiN

### 4. SUMMARY

Thermoreflectance methods by picosecond light pulse heating and by nanosecond light pulse heating have been developed under the same configuration as the laser flash method which is the standard method to measure the thermal diffusivity of bulk materials. Thermal diffusivity values across thin films were measured with small uncertainty. These high speed light pulse heating thermoreflectance methods can be observed the heat diffusion time across well-defined length of the film thickness under one-dimensional heat flow.

Using these light pulse heating methods, the thermal diffusivity of each layer of multilayered thin films and the boundary thermal resistance between the layers can be determined from the observed transient temperature curves based on the response function method.

The thermophysical properties of the transparent conductive films for flat panel display and the hard coating films were measured with the high speed light pulse heating thermoreflectance methods. The boundary thermal resistance between thin films and the thermal diffusivity of each layer of multilayer films constitute next generation phase-change optical disc were also measured with these methods. Reliable thermal design can be realized by heat conduction simulation using these reliable thermal diffusivity values of thin films and boundary thermal resistance values between thin films measured with the high speed light pulse heating thermoreflectance methods.

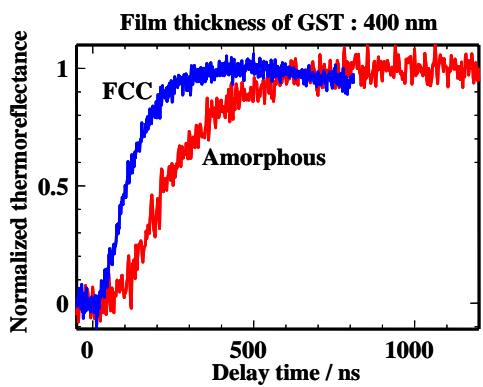

Figure 11  Thermoreflectance signals of Mo/$Ge_2Sb_2Te_5$/Mo three layered thin films in FCC structure and amorphous structure